\documentclass[a4paper,11pt,twoside]{article}
\usepackage{graphicx,asty}
\begin{document}
\title{\sc\huge ${\cal N}$-fold Supersymmetry\\ for
a Periodic Potential}
\author{\Large Hideaki Aoyama$^{\dagger,1}$, 
Masatoshi Sato$^{\ddagger,2}$, 
Toshiaki Tanaka$^{\dagger,3}$, \\[8pt]
\Large and Mariko Yamamoto$^\dagger$\footnote{Current address: 
Graduate School of Informatics, Kyoto University, Kyoto 606-8501, Japan}\\[8pt]
\sc $^\dagger$Faculty of Integrated Human Studies\\[4pt]
\sc Kyoto University, Kyoto 606-8501, Japan\\[7pt]
\sc $^\ddagger$The Institute for Solid State Physics\\[4pt]
\sc The University of Tokyo, Kashiwanoha 5-1-5, \\[4pt]
\sc Kashiwa-shi, Chiba 277-8581, Japan}
\footnotetext[1]{aoyama@phys.h.kyoto-u.ac.jp}
\footnotetext[2]{msato@issp.u-tokyo.ac.jp}
\footnotetext[3]{ttanaka@phys.h.kyoto-u.ac.jp}
\maketitle

\vspace{-10cm}
\rightline{KUCP-0169}
\vspace{9.5cm}
\thispagestyle{empty}
\def\nfsusy{${\cal N}$-fold supersymmetry}
\begin{abstract}
We report a new type of supersymmetry, ``\nfsusy", 
in one-dimen\break -sional quantum mechanics.
Its supercharges are $\cal N$-th order polynomials of 
momentum: It reduces to ordinary supersymmetry for ${\cal N}=1$, but
for other values of ${\cal N}$ the anticommutator of the supercharges
is not the ordinary Hamiltonian, but is a polynomial of the
Hamiltonian. 
(For this reason, the original Hamiltonian is referred to as
the ``Mother Hamiltonian".)
This supersymmetry shares some features with the ordinary variety,
the most notable of which is the non-renormalization theorem.
An \nfsusy\ was earlier found for a quartic potential whose
supersymmetry is spontaneously broken. Here
we report that it also holds for a periodic potential, 
albeit with somewhat different supercharges,
whose supersymmetry is {\it not} broken.\\[8pt]
\hspace{50pt}PACS numbers: 03.65.w, 03.65.Fd, 03.65.Ge, 11.30.Pd.\\
Keywords: Quantum mechanics, Supersymmetry,
Periodic potential, Quasi-solvable model.
\end{abstract}
\newpage

\section{Introduction}
Supersymmetric quantum mechanics has served as 
a testing and training ground for various concepts and ideas, 
for example the Witten index \cite{Witten2}, and
nonperturbative techniques before they are applied to
supersymmetric quantum field theories. 
It was recently found that when there is a quartic potential
it allows an extension to a new type of supersymmetry, which was dubbed
\nfsusy\ \cite{vv}.
This discovery was actually made through the non-renormalization
property found through the investigation of the 
nonperturbative properties of the theory by
the application of the valley method \cite{RR}--\cite{AKHOSW}:
In Refs.\cite{vv,AKOSW} the nonperturbative part of the energy spectrum
was calculated by the valley method, which, together with
an understanding of the Bogomolny's technique \cite{Bog} 
as the separation of the
purely nonperturbative piece from the perturbative piece, led to the
discovery of the disappearance of the leading Borel 
singularity of the perturbative series at some discrete
values of a parameter ($\cal N$) in the theory.
One such value (${\cal N}=1$) corresponded to
the case when the theory becomes supersymmetric and
the disappearance of the Borel singularity is explained 
by the fact that the ground state does not receive any
perturbative correction.
Thus it was speculated that at other integer values of ${\cal N}$,
a new symmetry similar to the supersymmetry may exist to
explain the non-renormalization properties, 
and in fact the authors of Ref.\cite{vv} succeeded in identifying it
and named it ``{\it \nfsusy}".

The scope of the \nfsusy, however, were limited:
While the ordinary supersymmetry allowed an arbitrary 
prepotential $W(q)$, under the $\cal N$-fold supercharges
defined in Ref.\cite{vv} the \nfsusy\ was possible only for
quadratic $W(q)$. 

In this letter we report a new \nfsusy\ for a periodic potential,
whose supercharges are different from the quadratic case.
In Section 2, we summarize the ordinary supersymmetry
and the \nfsusy\ for quadratic $W(q)$.
In Section 3, we prove a no-go theorem, which states that
under the same ${\cal N}$-fold supercharges as in Ref.\cite{vv}, 
only the quadratic $W(q)$ is possible.
The new \nfsusy\ for a periodic potential is discussed in Section 4.

\section{The quadratic case}
Let us first summarize the ordinary supersymmetry \cite{Witten,solo,cooper}
and set the notation.
Its two supercharges $Q$, $Q^\dagger$ and the Hamiltonian ${\bf H}$
satisfy the following algebra;
\begin{eqnarray}
&\{Q, Q\}=\{Q^\dagger, Q^\dagger\}=0, \label{eqn:qqdc}\\
&{\bf H}=\frac12\{Q^\dagger, Q\}, \label{eqn:hqqd}\\
&[{\bf H}, Q]=[{\bf H}, Q^\dagger]=0.\label{eqn:hqcom}
\end{eqnarray}
The actual representation is constructed by considering 
a particle with a one-dimensional bosonic coordinate (denoted by $q$)
and a fermionic coordinate ($\psi$), which
satisfy $\{\psi, \psi^\dagger\}=1$ and $\psi^2=\psi^{\dagger 2}=0$.
The supercharges are defined by the following:
\begin{equation}
Q=D^\dagger \psi, \quad Q^\dagger = D \psi^\dagger,
\end{equation}
where the operators $D^{(\dagger)}$ are defined by the following:
\begin{equation}
D=p-iW(q), \quad D^\dagger = p + iW(q),
\end{equation}
where $p=-i(d/dq)$ and $W(q)$ is an {\it arbitrary} real function of 
the coordinate $q$.
The Hamiltonian is given by the following;
\begin{equation}
{\bf H}=\frac12\left(p^2+W^2(q)\right)+
W'(q)\left(\psi^\dagger\psi-\frac12\right),
\end{equation}
In contrast to supersymmetric field theories, this
theory allows the introduction of a matrix representation for the fermionic
coordinates;
\begin{equation}
\psi=\left(\begin{array}{cc}0 & 0\\1 & 0\end{array}\right), \quad
\psi^\dagger=\left(\begin{array}{cc}0 & 1\\0 & 0\end{array}\right).
\end{equation}
In this matrix representation, the supercharges are written as follows:
\begin{equation}
Q=\left(\begin{array}{cc}0 & 0\\D^\dagger & 0\end{array}\right), \quad
Q^\dagger=\left(\begin{array}{cc}0 & D\\0 & 0\end{array}\right);
\label{eqn:qqd}
\end{equation}
while the Hamiltonian is
\begin{equation}
{\bf H}
=\left(\begin{array}{cc}H_+ & 0\\0 & H_-\end{array}\right),
\end{equation}
where
\begin{equation}
H_+=\frac12DD^\dagger, \quad H_-=\frac12D^\dagger D,
\label{eqn:h12dd}
\end{equation}
or in terms of $W(q)$,
\begin{equation}
H_{\pm}=\frac12 p^2+ \frac12\left(W^2(q)\pm W'(q)\right).
\label{eqn:hpm}
\end{equation}
In this matrix notation, the proof of the 
algebra (\ref{eqn:qqdc})--(\ref{eqn:hqcom})
is straightforward.  Specifically, the components of (\ref{eqn:hqcom}) 
are the following conjugate pair;
\begin{equation}
DH_-=H_+D, \quad H_-D^\dagger=D^\dagger H_+,
\end{equation}
which is a trivial relation derived from Eq.(\ref{eqn:h12dd}).
In the following, we use the matrix notation 
Eq.(\ref{eqn:qqd})--Eq.(\ref{eqn:hpm}).

\begin{figure}[ht]
\begin{center}
\includegraphics[width=9cm]{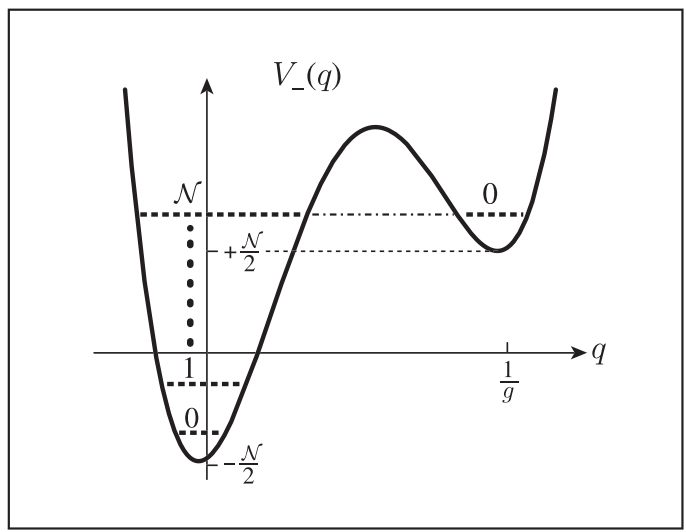}
\caption{The potentials of $H_{-\cal N}$ for $W(q)=q(1-gq)$.
The potential for $H_{+\cal N}$  is obtained by a reflection
$q\rightarrow 1/g-q$. The energy levels are denoted by
broken lines.}
\label{fig:quad}
\end{center}
\end{figure}

The ${\cal N}$-fold supercharges are defined by the following:
\begin{equation}
Q_{\cal N}=\left(\begin{array}{cc}0 & 0\\D^{\dagger {\cal N}}& 
0\end{array}\right), \quad
Q_{\cal N}^\dagger
=\left(\begin{array}{cc}0 & D^{\cal N}\\0 & 0\end{array}\right), 
\label{eqn:qqdn}
\end{equation}
and the Hamiltonian is;\footnote{The reader may note that
the notation for the Hamiltonians are different, 
a bit more streamlined than in Ref.\cite{vv}.}
\begin{equation}
{\bf H}_{\cal N}
=\left(\begin{array}{cc}H_{+{\cal N}} & 0\\0 & H_{-{\cal N}}\end{array}\right),
\label{eqn:hamin}
\end{equation}
where
\begin{equation}
H_{\pm{\cal N}}=
\frac12 p^2+ \frac12\left(W^2(q)\pm {\cal N}W'(q)\right).
\label{eqn:hpmn}
\end{equation}
The following is the ${\cal N}$-fold supersymmetric algebra:
\begin{eqnarray}
&\{Q_{\cal N}, Q_{\cal N}\}=\{Q_{\cal N}^\dagger, Q_{\cal N}^\dagger\}=0, \\
&[{\bf H}_{\cal N}, Q_{\cal N}]=[{\bf H}_{\cal N}, Q_{\cal N}^\dagger]=0,
\label{eqn:hqcomn}
\end{eqnarray}
which was proven for any integer ${\cal N}$
when $W(q)$ is a quadratic function of $q$.
(A proof of this is given in the next section.)
A convenient canonical form of $W(q)$ is the following:
\begin{equation}
W(q)=q(1-gq),
\label{eqn:wq}
\end{equation}
where the parameter $g$ is an analogue of the coupling constant.
For small ${\cal N}g^2$, $H_{\pm\cal N}$ have asymmetric double-well
potentials separated by the potential barrier of height of O($1/g^2)$
(see Fig.\ref{fig:quad}).

At $g=0$, the two wells become disjoint and two spectrum towers appear
at around $q=0$ and $q=1/g$.
For integer values of ${\cal N}$, the ${\cal N}$-th excited state
is degenerate, with the lowest state in the other tower.
This is the case when the \nfsusy\ exists and it protects the
degeneracy when the coupling constant $g$ is turned on.
The lowest ${\cal N}$-states that do not have the 
corresponding states in the other spectrum are called
``{\it Isolated States}", which are free from Borel singularities
due to the \nfsusy, and in fact the perturbative series for the
energy eigenvalues of these states have a finite convergence 
radius in the $g$-plane \cite{vv}.

The Hamiltonian ${\bf H}_{\cal N}$ is {\it not} equal to
the anticommutator $\frac12\{Q_{\cal N}^\dagger, Q_{\cal N}\}$
unless ${\cal N}=1$, which is the ordinary supersymmetric case.
This is evident from the fact that $Q_{\cal N}$ contains 
${\cal N}$-derivatives with respect to the coordinate $q$
and therefore $\frac12\{Q_{\cal N}^\dagger, Q_{\cal N}\}$ contains 
$2{\cal N}$-derivatives. The latter has, on the other hand,
a "family resemblance" to the Hamiltonian, 
and is thus called the ``{\it Mother Hamiltonian}";
\begin{equation}
{\cal H}_{\cal N}\equiv \frac12\{Q_{\cal N}^\dagger, Q_{\cal N}\} 
=\left(\begin{array}{cc}
\displaystyle\frac12D^{\cal N}D^{\dagger \cal N} & 0\\
0 & \displaystyle\frac12D^{\dagger \cal N} D^{\cal N}\end{array}\right).
\end{equation}
The following commutation relation is, of course, satisfied:
\begin{eqnarray}
[{\cal H}_{\cal N}, Q_{\cal N}]=[{\cal H}_{\cal N}, Q_{\cal N}^\dagger]=0.
\end{eqnarray}
It was conjectured that this Mother Hamiltonian ${\cal H}_{\cal N}$
is a polynomial of the ordinary Hamiltonian ${\bf H}_{\cal N}$;
\begin{equation}
{\cal H}_{\cal N}=\frac12\det{\bf M}_{\cal N}({\bf H}_{\cal N}),
\label{eqn:mh}
\end{equation}
where ${\bf M}_{\cal N}$ is a ${\cal N}\times{\cal N}$ matrix
obtained from the perturbative solution of the ${\cal N}$
isolated states \cite{vv}.  The identity (\ref{eqn:mh}) was proven for
${\cal N}=1,2,3$ by hand, and for values up to ${\cal N}=10$ by
the use of Mathematica.  

\section{A no-go theorem}
As noted before the \nfsusy\ for ${\cal N}=2,3,\cdots$
were found only for the particular $W(q)$ defined by Eq.(\ref{eqn:wq}),
while the ${\cal N}=1$ ordinary supersymmetry holds for any arbitrary
function $W(q)$.  Therefore it is most natural to explore what 
other kinds of $W(q)$ allow \nfsusy.
Furthermore, for the quadratic $W(q)$ the supersymmetry is spontaneously
broken. Therefore it is interesting to examine if \nfsusy\ exists
without spontaneous supersymmetry breaking.

It is possible to explore what kind of $W(q)$ is allowed
under the supercharges (\ref{eqn:qqdn}), but with the
following generalized Hamiltonian:
\begin{equation}
H_{\pm{\cal N}}=
\frac12 p^2+ \frac12\left(W^2(q)\pm f_{\cal N}^{(\pm)}(q)\right).
\label{eqn:hpmnrelaxed}
\end{equation}
The components of the desired commutation relation (\ref{eqn:hqcomn}) 
are the following:
\begin{equation}
D^{\cal N} H_{-\cal N}=H_{+\cal N} D^{\cal N},
\label{eqn:oldnfold}
\end{equation}
and its conjugate. 
In order to find the constraints for $W(q)$ and $f_{\cal N}^{(\pm)}(q)$
induced by the above relation, we will calculate the 
difference between the l.h.s.~and the r.h.s.~of the above:
\begin{eqnarray}
&&2(D^{\cal N} H_{-\cal N}-H_{+\cal N} D^{\cal N})\nonumber\\
&&\quad=D^{\cal N}(D^2+W'+2iWD-f_{\cal N}^{(-)})
-(D^2+W'+2iWD+f_{\cal N}^{(+)})D^{\cal N}\nonumber\\
&&\quad = D^{\cal N}W'-W'D^{\cal N} + 2i D^{\cal N}WD-2iWD^{{\cal N}+1}
-D^{\cal N}f_{\cal N}^{(-)}-f_{\cal N}^{(+)}D^{\cal N}
\end{eqnarray}
where we used a relation $p^2+W^2=D^2+W'+2iWD$. 
Since 
\begin{equation}
D=-i \,U^{-1}\partial \,U, \quad
U=e^{\int W(q)dq}
\end{equation}
where $\partial=\partial/\partial q$, we may transform all the $D$s
to $\partial\,$s by the $U$-transformation;
\begin{eqnarray}
&&i^{\cal N}U 2(D^{\cal N} H_{-\cal N}-H_{+\cal N} D^{\cal N}) U^{-1}
\nonumber\\
&&\qquad = \partial^{\cal N}W'-W'\partial^{\cal N} 
+ 2 \partial^{\cal N}W\partial-2W\partial^{{\cal N}+1}
-\partial^{\cal N}f_{\cal N}^{(-)}-f_{\cal N}^{(+)}\partial^{\cal N}
\end{eqnarray}
The next step is to move all the derivatives to the right and 
examine the coefficients of a given power of $\partial$.
It is straightforward to show that the coefficients
of $\partial^{{\cal N}+2}$ and $\partial^{{\cal N}+1}$ vanish identically.
The $\partial^{{\cal N}}$ term and $\partial^{{\cal N}-1}$ term
yields the following, respectively:
\begin{eqnarray}
&2{\cal N}W'-f_{\cal N}^{(+)}-f_{\cal N}^{(-)}=0\\
&{\cal N}W''-f_{\cal N}^{(-)\prime}=0
\end{eqnarray}
These lead to;
\begin{equation}
f_{\cal N}^{(-)}={\cal N}W'+C, \quad
f_{\cal N}^{(+)}={\cal N}W'-C, 
\end{equation}
where $C$ is an integration constant.  This reproduces the
original Hamiltonian (\ref{eqn:hamin}) with a meaningless constant $C$ added.
From the coefficient of $\partial^{{\cal N}-2}$ we find the following:
\begin{equation}
-\frac16{\cal N}({\cal N}-1)({\cal N}+1)W^{\prime\prime\prime}=0.
\end{equation}
The lower powers of $\partial$ have coefficients proportional to
the derivatives of $W$ of order four or more.
Therefore, we conclude that under the assumption of the 
form of the supercharges (\ref{eqn:qqdn}) and the Hamiltonian 
(\ref{eqn:hpmnrelaxed}), the commutation relation  (\ref{eqn:hqcomn}) holds
only for the following three cases:
(1) ${\cal N}=0$, the trivial case with arbitrary $W(q)$, 
(2) ${\cal N}=\pm 1$, the ordinary supersymmetry with arbitrary $W(q)$, 
and 
(3) $W^{\prime\prime\prime}=0$, the quadratic $W(q)$ found previously.

This proves that 
as long as the supercharges are of the form (\ref{eqn:qqdn}),
no new \nfsusy\ is possible.

\section{The periodic case}
We will now examine the following case:
\begin{equation}
W(q)=\frac1g\sin(gq),
\end{equation}
with periodicity $2\pi/g$. 
Ordinary supersymmetry is not broken since the region of
$q$ is finite, unlike the quadratic case.

\begin{figure}[ht]
\begin{center}
\includegraphics[width=9cm]{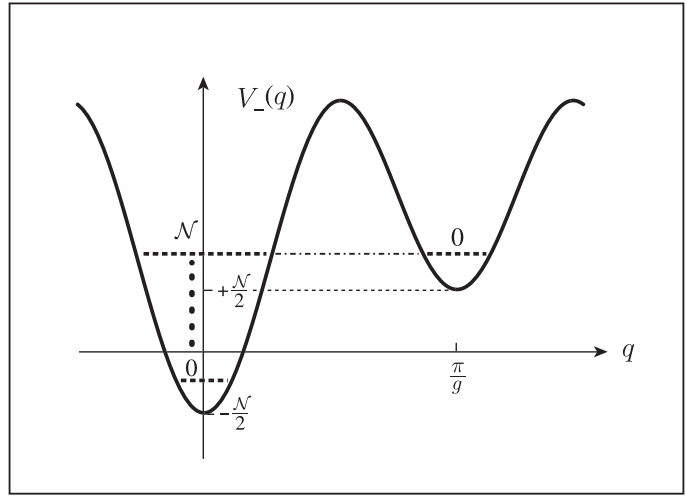}
\caption{The potential of $H_{-\cal N}$ for $W(q)=(1/g)\sin(gq)$
for $q \in [-\pi/(2g), 3\pi/(2g)]$.
The potential for $H_{+\cal N}$  is obtained by a reflection
$q\rightarrow \pi/(2g)-q$. The energy levels are denoted by
broken lines.}
\label{fig:peri}
\end{center}
\end{figure}

We define new
${\cal N}$-fold supercharges by the following:
\begin{eqnarray}
Q_{\cal N}=\left(\begin{array}{cc}0 & 0\\P_{\cal N}^\dagger& 
0\end{array}\right), &&\quad
Q_{\cal N}^\dagger
=\left(\begin{array}{cc}0 & P_{\cal N}\\0 & 0\end{array}\right), 
\label{eqn:qqdnperi}\\
P_{\cal N}\equiv &&\prod_{k=-{\cal M}}^{\cal M} (D+kg),
\end{eqnarray}
where ${\cal N}=2{\cal M}+1$.  Note that
the above product is always 
in the range $-{\cal M}, -{\cal M}+1, \cdots , {\cal M}-1, {\cal M}$
even for the half integer ${\cal M}$ (even ${\cal N}$).
The supercharges (\ref{eqn:qqdnperi}) contain ${\cal N}$-derivatives,
but are different from the form (\ref{eqn:qqdn}), and are 
free from the constraint of the no-go theorem proven in the
previous section.
The Hamiltonian is the same as in Eqs.(\ref{eqn:hamin}) and (\ref{eqn:hpmn}).
The potential of $H_{-\cal N}$ is illustrated in Fig.\ref{fig:peri}.

We first prove the commutation relation (\ref{eqn:hqcomn}), whose components
are; 
\begin{equation}
P_{\cal N}H_{-\cal N}=H_{+\cal N}P_{\cal N},
\label{eqn:periphn}
\end{equation}
and its conjugate.
As in the previous section, we transform the both sides of 
the above identity by the $U$-transformation.
The following is useful for this calculation:
\begin{eqnarray}
U P_{\cal N} U^{-1} &=& \prod_{k=-{\cal M}}^{\cal M} (-i\partial+kg),\\
U H_{-\cal N} U^{-1} &=&
-\frac12\partial^2 +\frac{e^{igq}}{2g}(-i\partial-{\cal M}g)
-\frac{e^{-igq}}{2g}(-i\partial+{\cal M}g),\label{eqn:hntrans}\\
U H_{+\cal N} U^{-1} &=& -\frac12\partial^2
+\frac{e^{igq}}{2g}\left(-i\partial+({\cal M}+1)g\right)
-\frac{e^{-igq}}{2g}\left(-i\partial-({\cal M}+1)g\right).
\end{eqnarray}
The l.h.s. of Eq.(\ref{eqn:periphn}) is then calculated as follows:
\begin{eqnarray}
U P_{\cal N}H_{-\cal N} U^{-1}
 &=& -\frac12\partial^2P_{\cal N}
+\frac{e^{igq}}{2g}
\prod_{k=-{\cal M}}^{\cal M} \left(-i\partial+(k+1)g\right)
(-i\partial-{\cal M}g)\nonumber\\
&& -\frac{e^{-igq}}{2g}
\prod_{k=-{\cal M}}^{\cal M} \left(-i\partial+(k-1)g\right)
(-i\partial+{\cal M}g),
\end{eqnarray}
which is identical to $U H_{+\cal N}P_{\cal N}U^{-1}$.
This completes the proof of (\ref{eqn:periphn}).

There is an alternative, more complex but more direct, 
proof of the identity (\ref{eqn:periphn}), 
which is similar to the proof of the quadratic case in Ref.\cite{vv}. 
We first note the following relation:
\begin{equation}
H_{+{\cal N}} D = D H_{+({\cal N}-2)}
+\frac{g}{4} ({\cal N}-1)\left[ -e^{igq}+e^{-igq}\right].
\end{equation}
Using this relation repeatedly, 
it is possible to show the following relation for any integer
${\cal N}'(=2{\cal K}+1)$ and ${\cal N}(=2{\cal M}+1)$ 
by the mathematical induction for ${\cal N}'$:
\begin{eqnarray}
H_{+\cal N} P_{\cal N'} &=& P_{\cal N'} H_{+({\cal N}-2{\cal N'})}
+\frac{g}{4}{\cal N'}({\cal N -N'})
P_{{\cal N'}-2} \left[ -(D-{\cal K}g)e^{igq}+(D+{\cal K}g)e^{-igq}\right].
\nonumber\\
\end{eqnarray}
This relation reproduces the desired commutation relation (\ref{eqn:periphn})
for ${\cal N}'={\cal N}$.

If we now turn to the isolated states and the Mother Hamiltonian next,
we see that in the ordinary \nfsusy, 
the algebra (\ref{eqn:oldnfold}) guarantees 
that any solution $\Psi_{-\cal N}$ 
of the Shr\"odinger equation of the Hamiltonian
$H_{-{\cal N}}$  is a solution 
of the Hamiltonian $H_{+{\cal N}}$ with the same eigenvalue,
{\it unless} it is eliminated by $D^{\cal N}$;
\begin{equation}
D^{\cal N}\Psi_{-\cal N}=0.
\label{eqn:elimi}
\end{equation}
This equation is satisfied by 
\begin{equation}
\Psi_{-\cal N}=f(q) \, e^{-\int W(q)dq},
\label{eqn:psifn}
\end{equation}
where $f(q)$ is a polynomial of order ${\cal N}-1$ or less.
The solution (\ref{eqn:psifn}) is not normalizable due to the $g q^2$-term
in $W(q)$, but it is so at any finite order of the 
perturbation expansion in $g$.
Thus the perturbative properties, especially the energy eigenvalues,
of the isolated states are determined completely by the \nfsusy.  
The fact that the solution (\ref{eqn:psifn}) is not normalizable
means that there are nonperturbative corrections on the
isolated states and the \nfsusy\ is spontaneously broken \cite{vv}.

In the current case, we have
\begin{equation}
P_{\cal N}\Psi_{-\cal N}=0,
\end{equation}
in place of (\ref{eqn:elimi}), which is satisfied by $f(q)$ of the form of
the truncated Fourier series;
\begin{equation}
f(q)=\sum_{k=-{\cal M}}^{\cal M}a_{k} \, e^{-ikgq}.
\end{equation}
By substituting the above into the Shr\"odinger equation
$H_{-\cal N}\Psi_{-\cal N}=E\Psi_{-\cal N}$ and using (\ref{eqn:hntrans}),
we find the following set of equations for the coefficients 
($a_{-\cal M}, \cdots a_{\cal M}$):
\begin{equation}
(k+1+{\cal M})a_{k+1}+(2E-k^2g^2)a_k-(k-1-{\cal M})a_{k-1}=0,
\end{equation}
for $k=-{\cal M}, \cdots {\cal M}$,
where $a_{{\cal M}+1}$ and $a_{-({\cal M}+1)}$ should be understood to
be zero. We rewrite the above to the matrix equation for the
vector ${\bf a}_{\cal N}=(a_{-\cal M}, \cdots, a_{\cal M})^{\rm T}$ as follows:
\begin{equation}
{\bf M}_{\cal N}(E)\cdot{\bf a}_{\cal N}=0.
\end{equation}
The energies of the isolated states are obtained from the following
condition that nontrivial ${\bf a}_{\cal N}$ are allowed as 
solutions of the above;
\begin{equation}
\det{\bf M}_{\cal N}(E)=0.
\label{eqn:yatta}
\end{equation}
It should be noted that unlike the quadratic case, the solutions
of Eq.(\ref{eqn:yatta}) are exact: Their wavefunctions are
normalizable and therefore
there are no nonperturbative corrections to the energy levels
obtained from Eq.(\ref{eqn:yatta}).  Since Eq.(\ref{eqn:yatta}) 
is a polynomial 
equation for $E$, its solutions have a 
finite convergence radius in $g$ and thus
is Borel-summable.  For example, at ${\cal N}=3$, Eq.(\ref{eqn:yatta})
yields the following;
\begin{equation}
\det{\bf M}_3(E)=(2E-g^2)(4(E^2-1)-2Eg^2)=0,
\end{equation}
whose solutions have infinite convergence radius.
It is interesting to note that
the {\it exact} energy of the first excited state is $E=g^2/2$,
which is only the first order correction, which is analogous to the
ABS anomaly.
It should be noted that these solutions for the 
isolated states come with specific boundary conditions;
periodic for odd ${\cal N}$ and anti-periodic for
even ${\cal N}$.  No exact solutions with other
twisted boundary conditions has been found so far.

We conjecture that the Mother Hamiltonian ${\cal H}_{\cal N}$
defined by 
\begin{equation}
{\cal H}_{\cal N}\equiv \frac12\{Q_{\cal N}^\dagger, Q_{\cal N}\}, 
\end{equation}
with the supercharges (\ref{eqn:qqdnperi})
is again given by the following polynomial of the usual
Hamiltonian;
\begin{equation}
{\cal H}_{\cal N}=\frac12\det{\bf M}_{\cal N}({\bf H}_{\cal N}).
\label{eqn:mh2}
\end{equation}
We have proven the above identity directly for ${\cal N}=3$
by the use of Mathematica.  

We have found some further extensions of the 
${\cal N}$-fold supersymmetry are possible, including
cubic and exponential $W(q)$s.  This and further
general discussions will be published in the near future.

\vspace*{12pt}
\centerline{\large\bf Acknowledgment}
\vspace*{12pt}
The authors would like to thank Dr.~Hisashi Kikuchi
(Ohu University, Japan) for discussions and Dr.~John Constable
(Magdalene College, U.K.) for reading the manuscript.
H.~Aoyama's work was supported in part by the Grant-in-Aid 
for Scientific Research No.10640259.
T.~Tanaka's work was supported in part by a JSPS research fellowship.


\begin{thebibliography}{99}
\def\J#1#2#3#4{{\sl #1} {\bf #2} (#3) #4}
\def\PL{Phys. Lett.}
\def\NP{Nucl. Phys.}
\bibitem{Witten2} E. Witten, \J{\NP}{B202}{1982}{253}.
\bibitem{vv} 
H. Aoyama, H. Kikuchi, I. Okouchi, M. Sato, and   S. Wada, 
\J{\NP}{B553}{1999}{644}.
\bibitem{RR}
D. J. Rowe and A. Ryman, \J{J. Math. Phys.}{23}{1982}{732}.
\bibitem{BY2}
I. I. Balitsky and A. V. Yung, \J{\PL}{B168}{1986}{13}.
\bibitem{sil}
P. G. Silvetrov, \J{Sov. J. Nucl. Phys.}{51}{1990}{1121}.
\bibitem{AK}
H. Aoyama and H. Kikuchi, \J{\NP}{B369}{1992}{219}.
\bibitem{AW}
H. Aoyama and S. Wada, \J{\PL}{B349}{1995}{279}.
\bibitem{HS} T. Harano and M. Sato, {\sl hep-ph}/9703457. 
\bibitem{AKHSW} 
H. Aoyama, H. Kikuchi, T. Harano, M. Sato, and S. Wada, 
\J{Phys. Rev. Lett.}{79}{1997}{4052}.
\bibitem{AKHOSW}H. Aoyama, H. Kikuchi, T. Harano, I. Okouchi, M. Sato, S. Wada,
\J{Prog. Theor. Phys. Supplement}{127}{1997}{1}.
\bibitem{AKOSW} H. Aoyama, H. Kikuchi, I. Okouchi, M. Sato, and
  S. Wada, \J{\PL}{B424}{1998}{93}.
\bibitem{Bog}E. B. Bogomolny, \J{\PL}{B91}{1980}{431}.
\bibitem{Witten} E. Witten, \J{\NP}{B188}{1981}{513}.
\bibitem{solo} P. Solomonson and J. W. Van Holten,
\J{\NP}{B196}{1982}{509}.
\bibitem{cooper} F. Cooper, A. Khare, and U. Sukhatme,
\J{Phys. Rep.}{251}{1995}{267}.
\end{thebibliography}
\end{document}